\documentclass[floatfix, noshowpacs, preprintnumbers, twocolumn, amsmath, amssymb, aps, prb, superscriptaddress]{revtex4-2}

\usepackage{graphicx, xcolor}
\usepackage{soul}
\usepackage{mathrsfs}
\usepackage{float}
\usepackage{subfloat}
 
\usepackage{amsmath,amscd,amsbsy,amssymb,latexsym,url,bm,amsthm}
\usepackage[vlined,boxed,commentsnumbered,linesnumbered,ruled]{algorithm2e}

\newcommand{\RNum}[1]{\uppercase\expandafter{\romannumeral #1\relax}}

\newcommand{\e}{\mathrm{e}}
\renewcommand{\i}{\mathrm{i}}

\begin{document}

\title{Collision-assisted information scrambling on a configurable photonic chip} 

\author{Xiao-Wen Shang}
\altaffiliation{These authors contributed equally to this work.}
\affiliation{Center for Integrated Quantum Information Technologies (IQIT), School of Physics and Astronomy and State Key Laboratory of Photonics and Communications, Shanghai Jiao Tong University, Shanghai 200240, China}
\affiliation{Hefei National Laboratory, Hefei 230088, China}

\author{Shu-Yi Liang}
\altaffiliation{These authors contributed equally to this work.}
\affiliation{School of Physics and Astronomy, Shanghai Jiao Tong University, Shanghai 200240, China}

\author{Guan-Ju Yan}
\altaffiliation{These authors contributed equally to this work.}
\affiliation{School of Physics and Astronomy, Shanghai Jiao Tong University, Shanghai 200240, China}

\author{Xin-Yang Jiang}
\affiliation{School of Physics and Astronomy, Shanghai Jiao Tong University, Shanghai 200240, China}

\author{Zi-Ming Yin}
\affiliation{School of Physics and Astronomy, Shanghai Jiao Tong University, Shanghai 200240, China}

\author{Hao Tang}
\email{htang2015@sjtu.edu.cn}
\affiliation{Center for Integrated Quantum Information Technologies (IQIT), School of Physics and Astronomy and State Key Laboratory of Photonics and Communications, Shanghai Jiao Tong University, Shanghai 200240, China}
\affiliation{Hefei National Laboratory, Hefei 230088, China}

\author{Jian-Peng Dou}
\affiliation{Center for Integrated Quantum Information Technologies (IQIT), School of Physics and Astronomy and State Key Laboratory of Photonics and Communications, Shanghai Jiao Tong University, Shanghai 200240, China}
\affiliation{Hefei National Laboratory, Hefei 230088, China}

\author{Ze-Kun Jiang}
\affiliation{Center for Integrated Quantum Information Technologies (IQIT), School of Physics and Astronomy and State Key Laboratory of Photonics and Communications, Shanghai Jiao Tong University, Shanghai 200240, China}
\affiliation{Hefei National Laboratory, Hefei 230088, China}

\author{Yu-Quan Peng}
\affiliation{TuringQ Co., Ltd., Shanghai 200240, China}

\author{Xian-Min Jin}
\email{xianmin.jin@sjtu.edu.cn}
\affiliation{Center for Integrated Quantum Information Technologies (IQIT), School of Physics and Astronomy and State Key Laboratory of Photonics and Communications, Shanghai Jiao Tong University, Shanghai 200240, China}
\affiliation{Hefei National Laboratory, Hefei 230088, China}
\affiliation{TuringQ Co., Ltd., Shanghai 200240, China}
\affiliation{Chip Hub for Integrated Photonics Xplore (CHIPX), Shanghai Jiao Tong University, Wuxi 214000, China}


\begin{abstract}
Quantum interference and entanglement are in the core of quantum computations. 
The fast spread of information in the quantum circuit helps to mitigate the circuit depth. 
Although the information scrambling in the closed systems has been proposed and tested in the digital circuits, how to measure the evolution of quantum correlations between systems and environments remains a delicate and open question. 
Here, we propose a photonic circuit to investigate the information scrambling in an open quantum system by implementing the collision model with cascaded Mach-Zehnder interferometers. 
We numerically simulate the photon propagation and find that the tripartite mutual information strongly depends on the system-environment and environment-environment interactions. 
We further reduce the number of observables and the number of shots required to reconstruct the density matrix by designing an enhanced compressed sensing. 
Our results provide a reconfigurable photonic platform for simulating open quantum systems and pave the way for exploring controllable dissipation and non-Markovianity in discrete-variable photonic computing. 
\end{abstract}

\maketitle

\emph{Introduction.}--Open quantum systems, where non-classical correlations take place within systems and external environments~\cite{von1950theory}, have been found rich in dynamical properties and have a significant impact on the field of quantum information, e.g., entanglement generation~\cite{diehl2008quantum,krauter2011entanglement,cho2011optical}, quantum computation~\cite{verstraete2009quantum} and quantum metrology~\cite{giovannetti2011advances,giovannetti2006quantum}. Although a wealth of theoretical approaches have been proposed, including Lindblad master equation~\cite{gorini1976completely,lindblad1976generators}, quantum trajectories~\cite{carmichael1993quantum} and quantum jumps~\cite{dum1992monte}, one crucial but not yet fully addressed issue is the flow of information in the open quantum system. An intuitive categorization is between Markovian and non-Markovian processes, where the former has no memory effect and the latter does \cite{Breuer2016}. 

Assessing the non-local diffusion of information requires delicate considerations, since one can still obtain the full range of information after incorporating and observing the environmental degrees of freedom. If the reservoir has a degree of freedom comparable to the system, quantum states are still prone to reconfiguration by local measurements, even if all the information has been transferred out of the system. Thus, the loss of information is not naively determined by the demarcation between the system and the environment. In closed systems, two indicators--out-of-time-order correlation (OTOC)~\cite{Swingle2016,Bohrdt2017,Campisi2017,Fan2017,Swingle2017,Swingle2018,Gonzalez2019,Landsman2019,Yoshida2019,Joshi2020Quantum,Mi2021} and tripartite mutual information (TMI)~\cite{Shen2020,Lin2021,Jafferis2022,Zhu2022}--have been proposed to measure whether the local information can be extracted by the local operations after evolution, \textit{i.e.}, whether the information has been scrambled~\cite{Hosur2016,Bhattacharyya2022}. To date, both OTOC and TMI have been experimentally applied to determine information scrambling (IS) in multiple platforms, e.g., ion traps~\cite{Joshi2020Quantum}, superconducting circuits~\cite{Blok2021,Mi2021,Jafferis2022,Zhu2022} and neutral-atom arrays~\cite{Bluvstein2024}. 

In open quantum systems, scrambled information has also been noticed recently~\cite{Li2020,Li2022}, especially the distinction with decoherence~\cite{Han2022}. However, experimentally probing IS in open quantum systems has long been a demanding task given the need of precise modulation of system-environment interactions and large-scale quantum state tomography to reconstruct the multipart density matrix within a coherence time. 

\begin{figure*}[t!]
\includegraphics[width=0.98\textwidth]{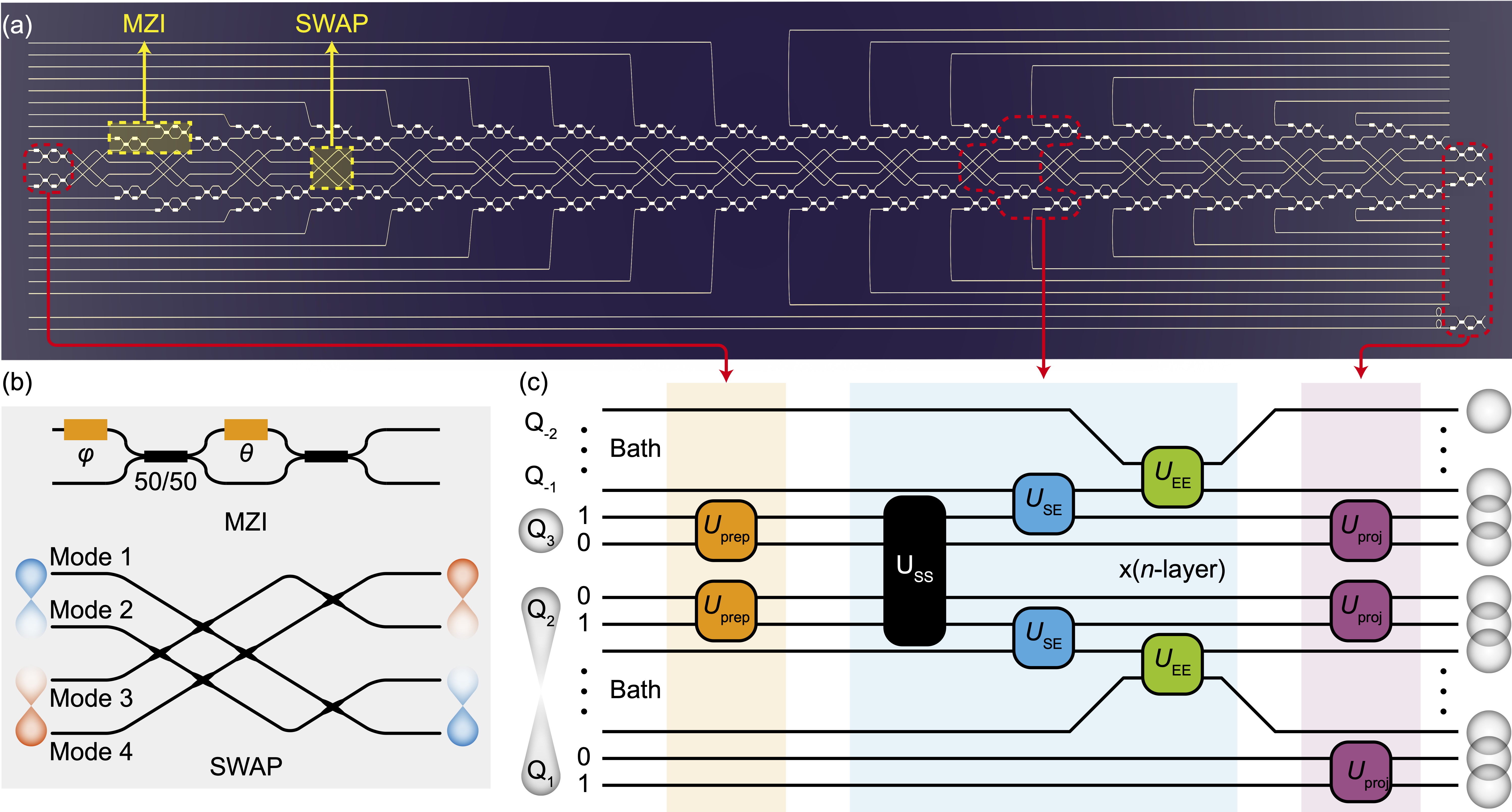}
\caption{\textbf{Circuit structure of simulating information scrambling in the open quantum system.} (a) The schematic diagram of the photonic integrated chip based on the quantum collision model. Single-mode waveguides are represented by gold lines. (b) Details of the Mach-Zehnder interferometer (MZI) and SWAP gate. Two phase shifters (yellow blocks) and two multimode interferometers used as two 50:50 beam splitters (black blocks) are included in the MZI. In the SWAP gate, six crosses are used to route photons. The state components on Modes 1 and 2 are respectively routed to Modes 4 and 3. (c) The schematic diagram of the gate circuit. The process is divided into three blocks, \textit{i.e.}, state preparation, collision and quantum state tomography. The collision block is repeated for $n$ times in the circuit. $Q_1$ is an ancilla qubit entangled with system qubit $Q_2$. $Q_3$ is the other system qubit. If photons are injected into the environment modes, the ambient qubits are encoded on $Q_{-1}$, $Q_{-2}$, \textit{etc.} }
\label{fig:Fig1}
\end{figure*}

Photons facilitate a promising platform for presenting open quantum systems benefiting from weak decoherence and high tunability~\cite{Aspuru-Guzik2012}. Here, we propose a specialized photonic circuit, where the dual rail~\cite{Ralph2002} and qumode are both used for encoding information, to investigate IS in open quantum systems. We decompose the process into a series of collisions between particles, \textit{i.e.}, apply a collision model~\cite{Ciccarello2013,McCloskey2014,Ciccarello2017,Jin2018,Cuevas2019,Rodrigues2019,Garcia2020,Cattaneo2021,Ciccarello2022,Cusumano2022,Li2024} for stroboscopic monitoring of the quantum state. 
Furthermore, we explore the technique of compressed sensing~\cite{gross2010quantum} and significantly reduce the number of observables and samples needed to be measured. It is found that the strength of the system-environment interaction strongly influences the speed and intensity of information scrambling. The effect of interactions between two environmental particles on non-Markovianity and TMI presents a similar picture. The proposed scheme provides a practical method for photonic simulation of many-body interactions in large-scale open quantum systems, expanding the scenario of quantum simulation.

\emph{Collision model and photonic circuit.}--We consider a two-particle system respectively interacting with two environments containing multi-particles. The interaction is realized by $n$ collision cycles. Each collision cycle is decomposed into three steps. In the first step, two system particles collide with each other (SS collision), in which the two particles exchange their quantum states. In the second step, the system particles collide with the environment particles separately (SE collision). In the third step, the environment modes that were collided with the system in the previous step are collided with two fresh vacuum environment modes (EE collision). 


\begin{figure*}[t!]
\includegraphics[width=0.98\textwidth]{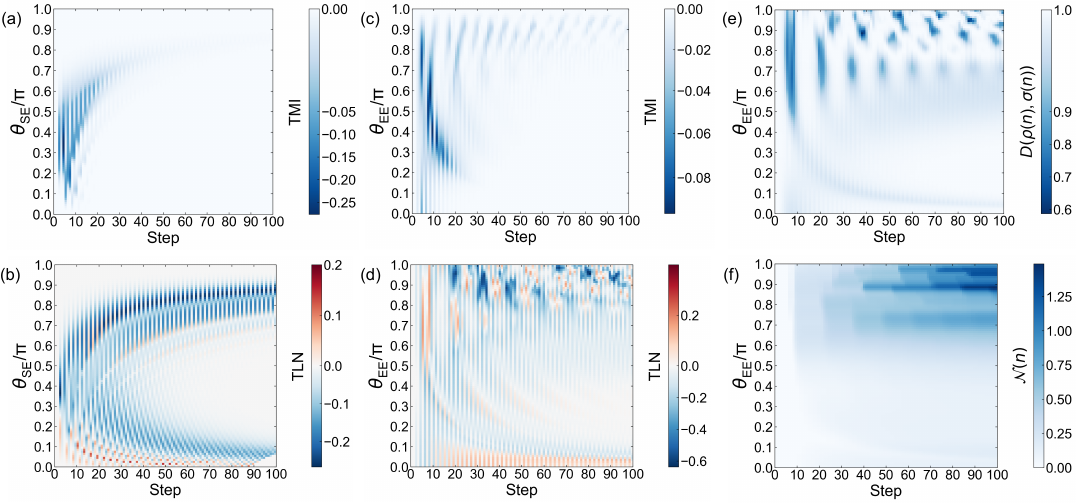}
\caption{\textbf{The evolution of TMI, TLN, trace distance and degree of non-Markovianity for continuous variation in the strength of SE or EE interactions.} (a) and (b) respectively illustrate the evolution of TMI and TLN with varying $\theta_{\mathrm{SE}}$, where the EE interaction is characterized by $\theta_{\mathrm{EE}}=3\pi/5$. (c), (d), (e) and (f) respectively display the evolution of TMI, TLN, trace distance $D(\rho(n),\sigma(n))$ and the degree of non-Markovianity $\mathcal{N}(n)$ with varying $\theta_{\mathrm{EE}}$, where the SE interaction is characterized by $\theta_{\mathrm{SE}}=\pi/4$.}
\label{fig:Fig2}
\end{figure*}

The on-chip photonic circuit used to interpret the information scrambling in the open quantum system is sketched in Fig.~\ref{fig:Fig1} (a). The photons are injected from the left ports, output at the right ports, and finally detected off-chip. Two qubits, $Q_1$ and $Q_2$, are used as system particles. We apply the dual-rail encoding~\cite{Ralph2002} to the system particles, \textit{i.e.}, each qubit is encoded onto two single mode waveguides, into which a single photon is injected. The position of the photon in the two waveguides indicates the qubit state. To calculate the information scrambling, an ancilla qubit $Q_{\mathrm a}$ is in a maximally entangled state with $Q_2$. The initial state can be expressed as
\begin{equation}
	|\Psi\rangle = \frac1{\sqrt{2}}|0\rangle_{-1}\otimes\left(|00\rangle_{\mathrm{1}2}+|11\rangle_{\mathrm{1}2}\right)\otimes|0\rangle_{3},
\end{equation}
where the state $|0\rangle_{-1}$ denotes a photon initially injected in the environment. 

The basic components of the circuit are the Mach-Zehnder interferometer (MZI) and SWAP gate, as shown in Fig.~\ref{fig:Fig1} (b). In an MZI, two phase shifters are used to change the phase of the photon by $\varphi$ (outer phase shifter) and $\theta$ (inner phase shifter), while two multimode interferometers are used as two 50:50 beam splitters. The MZI can achieve an arbitrary two-by-two unitary transformation of the two modes. The transformation matrix $U_{\mathrm{MZI}}$ reads
\begin{equation}
	U_{\mathrm{MZI}}(\theta, \varphi) = \i \e^{\i\theta/2}
	\begin{pmatrix}
		\e^{\i\varphi} \sin\left(\theta/2\right) &  \cos\left(\theta/2\right) \\
		\e^{\i\varphi} \cos\left(\theta/2\right) & -\sin\left(\theta/2\right) \\
	\end{pmatrix}.
\end{equation}

Arbitrary initial states of the system particles can be prepared by two MZIs that are separately implemented on $Q_1$ and $Q_2$. 
SS collision is implemented by a SWAP gate, leading to the transformation of the system state $|\psi\rangle_1|\phi\rangle_2$ to $|\phi\rangle_1|\psi\rangle_2$. This is a simulation of that two particles exchange their inner states by a collision. Considering that excessive MZIs spike the difficulty of characterization, the SWAP gate simulates the SS collision in a straightforward way. 
SE collision is realized by an MZI, which is applied to the Mode 1 of $Q_1$ ($Q_2$) and an environment mode. If $Q_1$ ($Q_2$) is in the state $|0\rangle$ (ground state), the SE collision does not affect the state of the system particle. Otherwise, the energy of the system particle partially flows to the environment.
EE collision is also realized by an MZI but with different modes. The `old' environment mode, which interplayed with the system mode in the previous step, interacts with a `fresh' vacuum environment mode by an MZI. By tuning the splitting ratio of the MZI, we can control the propagation direction of the photon which carries the information of the system. Thus, a Markovian (non-Markovian) process can be achieved by tuning $\theta$ of the MZI in the EE collision (see Fig.~\ref{fig:Fig1} (d)). 

The density matrix of the system and ancilla qubits $\hat{\rho}_{123}$ after the collisions can be theoretically reconstructed by the technology of quantum state tomography. 
Prior to the measurement, three MZIs are utilized to rotate the basis vectors of the projection measurement to the eigenstates of Pauli operators $\{\sigma_0,\sigma_1,\sigma_2,\sigma_3\}$. For each qubit, two single-photon detectors probe the probability of a photon being in each waveguide, enabling the Pauli-based measurements performed on each qubit. With the expectations of the Pauli operators, the density matrix can be calculated from the relationship ${\rho}_{123}=1/2^3\sum_{k_1,k_2,k_3=0}^3\left[\mathrm{Tr}({\sigma}_{k_1}{\sigma}_{k_2}{\sigma}_{k_3}\rho_{123}){\sigma}_{k_1}{\sigma}_{k_2}{\sigma}_{k_3}\right]$.

\emph{System-environment interaction and environment-environment interaction.}--As illustrated in Fig.~\ref{fig:Fig1}, the parameter $\theta\in[0,\pi]$ determines the reflectivity of an MZI. We use $\theta_{\mathrm{SE}}$ and $\theta_{\mathrm{EE}}$ to characterize the probability that a photon continues to propagate along the waveguide where it was before the collision after SE and EE collisions, respectively. To capture the diffusion of quantum correlations in the many-body quantum system, we simulate the TMI after each step of collisions. The TMI of the system and ancilla qubits can be expressed by
\begin{equation}\label{I3}
	\begin{aligned}
    I_3(Q_1:Q_2:Q_3) &= I_2(Q_1:Q_2)+I_2(Q_1:Q_3)\\
    &- I_2(Q_1:Q_2Q_3),
    \end{aligned}
\end{equation}
where $I_2(Q_1:Q_X)$ is the bipartite mutual information (BMI) of the ancilla qubit and the qubit $Q_X$ which depends on the von Neumann entropy $S\left(\rho\right)=-\mathrm{tr}\left(\rho\log\rho\right)$ by
\begin{equation}\label{I2}
    I_2(Q_1:Q_X)=S\left(\rho_1\right)+S\left(\rho_X\right)-S\left(\rho_{1X}\right).
\end{equation}
Combining Eq.~\ref{I3} and Eq.~\ref{I2} reveals that a negative TMI indicates that information about the whole quantum system cannot be obtained by measuring the subsystems, implying that information scrambling has occurred.

The effect of system and environment interactions on the information scrambling is illustrated in Fig.~\ref{fig:Fig2} (a), where the evolution of TMI with the number of collision steps for varying $\theta_{\mathrm{SE}}$ is plotted and mapped to the colorbar directly below the image. To demonstrate a general situation, a moderate strength of interaction between environmental particles is applied by setting $\theta_{\mathrm{EE}}=\pi/4$. In the range of $[0,\pi]$, a large $\theta_{\mathrm{SE}}$ realizes a large probability of the photon being in the original system state, providing a weak interaction between the system and the environment particles. It can be seen that the weaker the strength of the interaction between the system and the environment, the more collisions are needed for information scrambling to occur. 

However, it is found that the negative TMI can emerge even in the absence of information scrambling due to decoherence in open quantum systems~\cite{Han2022}. In this case, a more reliable indicator of scrambling is the tripartite logarithmic negativity (TLN) defined by~\cite{Han2022,Kudler-Flam2020}
\begin{equation}
\begin{aligned}
	N_3(Q_1:Q_2:Q_3) &= N_2(Q_1:Q_2)+N_2(Q_1:Q_3)\\
	&-N_2(Q_1:Q_2Q_3),
\end{aligned}
\end{equation}
where
\begin{equation}
	N_2(Q_1:Q_X)=\log\left(\left|\left|\rho_{1X}^{\mathrm{T}_X}\right|\right|_1\right)
\end{equation}
is the bipartite logarithmic negativity (BLN) and $\rho_{1X}^{\mathrm{T}_X}$ is the partial transpose of a density matrix with respect to the subsystem $Q_X$. The evolution of the TLN is drawn in Fig.~\ref{fig:Fig2} (b). 
A comparison between Figs.~\ref{fig:Fig2} (a) and (b) reveals distinct behaviors in the range of $\theta_{\text{SE}}\in [0.1\pi,0.3\pi]$ and $\text{step}\in[20,100]$. While in this region there is TLN negativity which is related to the decoherence effects that deteriorate the initial entanglement, there is barely no TMI negativity.  This suggests that the emergence of TMI negativity is primarily driven by scrambling-induced information delocalization, rather than by decoherence effects. 
Besides the collisions between the system and the environment, the interactions between environmental particles also influence the TMI and TLN, as displayed in Fig.~\ref{fig:Fig2} (c) and (d).


Since it is always environmental Mode 1 that interferes with the system mode, the information passed to the environmental mode in the last collision cannot flow back to the system if the transmissivity of the MZI composed of environmental Mode 1 and Mode 2 is 1, \textit{i.e.}, if $\theta_{\mathrm{EE}}=0$. The results presented in Fig.~\ref{fig:Fig2} (e) are consistent with the above analysis, in which we use the trace distance to describe the non-Markovianity~\cite{Liu2011}, \textit{i.e.}, 
\begin{equation}\label{D}
    D(\rho(n),\sigma(n)) = \frac{1}{2}\|\rho(n)-\sigma(n)\|_1,
\end{equation}
where $\|\cdot\|_1$ is the trace norm and $n$ is the collision step. Before evolution, the trace distance of two quantum states $\rho(0)$ and $\sigma(0)$ is 1. During the evolution, an increase in the trace distance indicates an enhancement of the distinguishability of the two states. Given that the distinguishability dissipates in a Markovian processes but recovers in non-Markovian processes~\cite{Liu2011}, the trend of $D(\rho(n),\sigma(n))$ with $n$ exhibits the non-Markovian properties in our open system. As shown in Fig.~\ref{fig:Fig2} (e), when $\theta_{\mathrm{EE}}>0$, and especially when $\theta_{\mathrm{EE}}>0.4$, there is a significant growth process in the trace distance. Comparing Figs.~\ref{fig:Fig2} (c), (d) and (e), the trends of non-Markovianity and information scrambling with respect to EE interaction strength are similar. Further, we calculate the degree of non-Markovianity $\mathcal{N}(n)$ by summarizing the trace distance over all increasing intervals, \textit{i.e.},
\begin{equation}
    \mathcal{N}(n)=\sum_{\substack{k=1,\\D(k+1)>D(k)}}^{n-1}\left[D(k+1)-D(k)\right].
\end{equation}
The results are illustrated in Fig.~\ref{fig:Fig2} (f). It can be seen that as $\theta_{\mathrm{EE}}$ increases, the strengthening of the interaction leads to an improvement in non-Markovianity.

\begin{figure}[!htbp]
\includegraphics[width=1\linewidth]{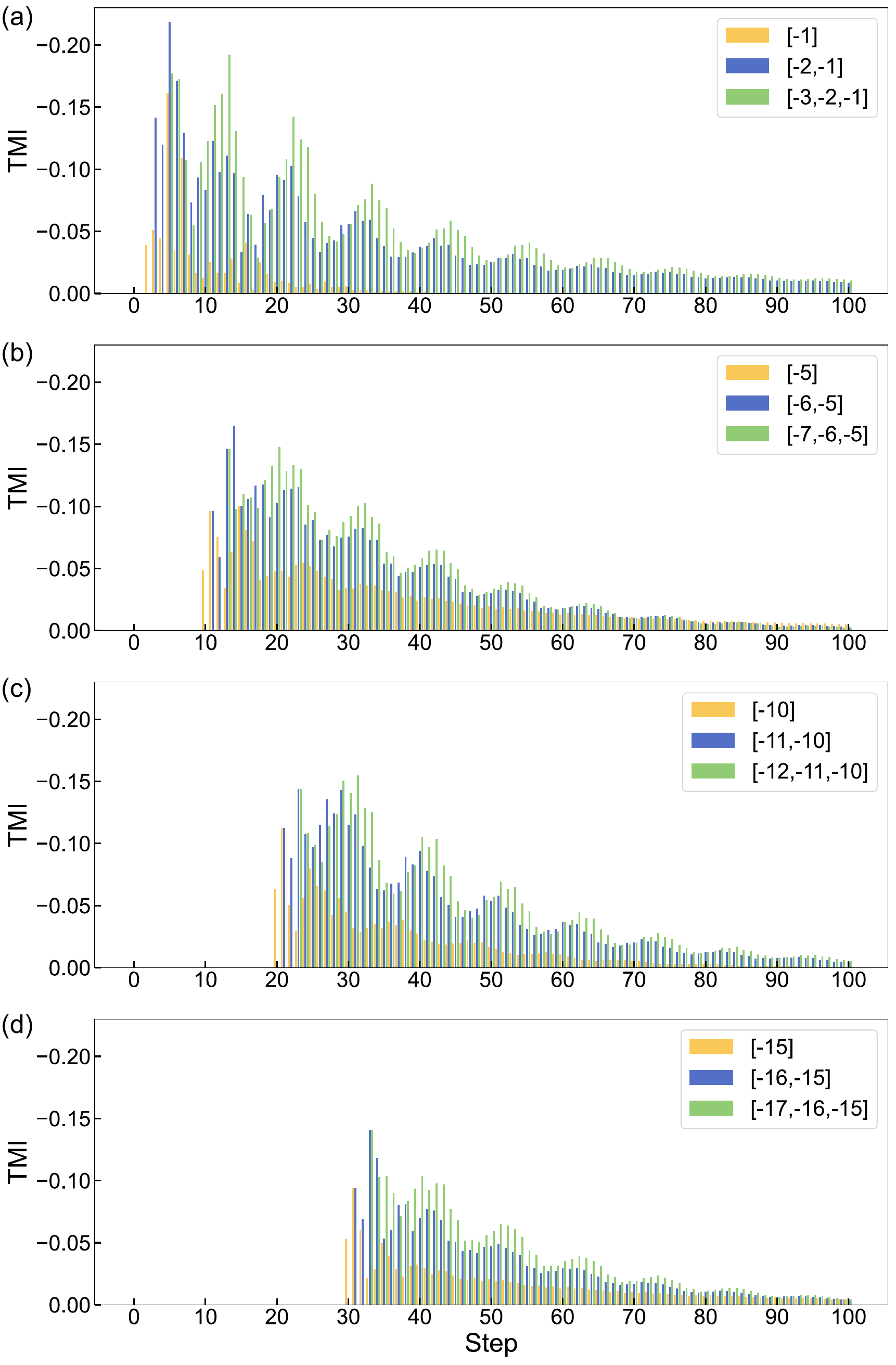}
\caption{\textbf{Effect of the number and position of ambient photons on TMI.} In (a), (b), (c) and (d), the injected ambient photons to $Q_3$ are initially at distances of $0$, $4$, $9$ and $14$ dual-rail modes, respectively. TMIs with one, two and three ambient photons are separately indicated by yellow, blue and green bars. The relative position of the ambient multi-photons is labeled in the upper right corner of the diagram. In all cases, the SE and EE interactions are characterized by $\theta_{\mathrm{SE}}=\theta_{\mathrm{EE}}=\pi/2$. Note that the positive direction of the vertical axis is downward.}
\label{fig:Fig3}
\end{figure}

\emph{Effect of ambient photons.}--In the above simulations, the injection of the photons is constrained to the system and ancillary system. It is also found that the presence of photons in the environment, \textit{i.e.}, the excitation of ambient qubits, can have some other effect on the information scrambling. The results are shown in Fig.~\ref{fig:Fig3}, where the positions of the excited ambient qubits $Q_p$ (see Fig.~\ref{fig:Fig1} (c)) are labeled in the upper right corner of each subgraph with $p$. The absolute value of TMI increases rapidly when environmental photons interact with the system and then slowly decays to zero. In each subgraph, one, two and three ambient qubits are excited at similar locations respectively. It is found that encoding information on more qubits in the environment that are involved in evolution leads to stronger information scramblings (more negative TMIs). Comparing Figs.~\ref{fig:Fig3} (a), (b), (c) and (d), the information scrambling occurs later as the ambient qubits are farther away from the system.


\begin{figure}[!htbp]
\includegraphics[width=0.9\linewidth]{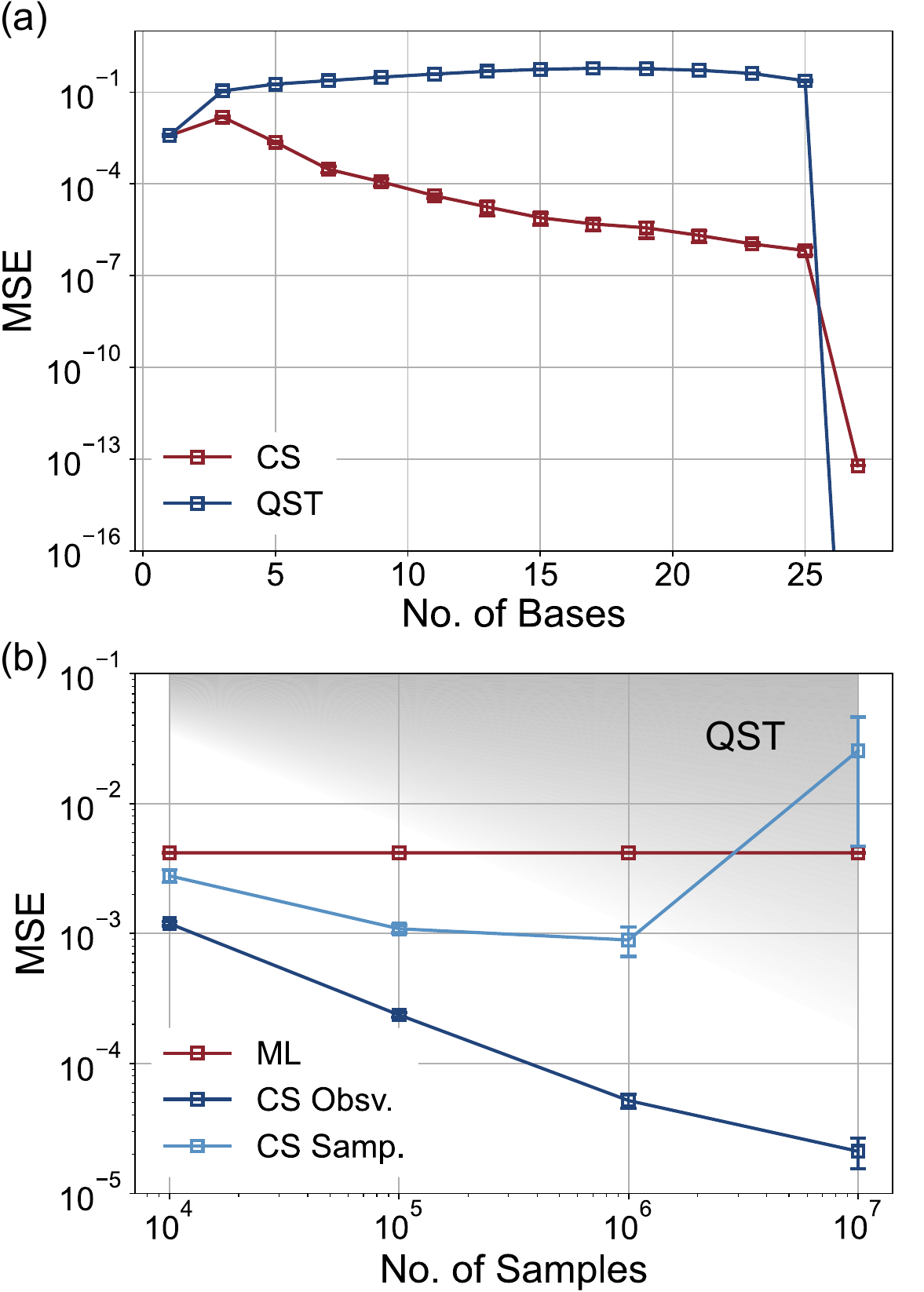}
\caption{\textbf{Precision of evaluating TMI by various methods.} (a) MSE of TMI evaluation by conducting CS and QST as a function of number of measurement bases. The bases are chosen randomly from all 3-qubit Pauli bases. For QST, expectations of inaccessible Pauli observables are set to $0$. (b) MSE of TMI evaluation by conducting ML, CS Obsv. and CS Samp. as a function of number of photon samplings. All samples are uniformly distributed to $14$ (half) 3-qubit Pauli bases. The regime wherever MSE is larger than that obtained by QST conducted on full $27$ Pauli bases is shaded. The ML model is trained on a separate set of 3-qubit states. All MSE are averaged over the results of total 1775 states generated with uniformly distributed $\theta_{\mathrm{SE}}$ and $\theta_{\mathrm{EE}}$. The error bars indicate standard deviation.
}
\label{fig:Fig4}
\end{figure}

\emph{Compressed sensing for TMI.}--Typically, the evaluation of TMI depends on the whole density matrix, which can be reconstructed by tomography~\cite{McGinley2022}. However, QST is susceptible to sampling noises and requires a large measurement overhead. Based on significantly lower number of bases and samples, quantum compressed sensing provides an alternative method~\cite{gross2010quantum}. Despite the fact that compressed sensing is meant for nearly-pure states, there have been applications of compressed sensing without such assumptions of purity or low rank~\cite{riofrio2017experimental,steffens2017experimentally}.

Original compressed sensing is conducted as:
\begin{equation}
    \operatorname*{min}\left|\left|\sigma\right|\right|_{\mathrm{Tr}}\,\,\mathrm{s.t.}\,\,\mathrm{Tr}\left(\sigma\right)=1,\,\,\mathrm{Tr}\left(\sigma P_k\right)=\mathrm{Tr}\left(\rho P_k\right),
\end{equation}
where $P_k$ traverses all possible direct products of $n$-qubit Pauli matrices, $\mathrm{Tr}\left(\rho P_k\right)$ is the expectation of observable $P_k$, and the solution $\sigma$ is a reconstruction of density matrix. Experimentally, expectations of Pauli observables that commute can be evaluated simultaneously from the same measurement basis. 

We compare the performances of CS and ordinary QST by calculating the mean-square error (MSE), as sketched in Fig.~\ref{fig:Fig4} (a). 
The precision of compressed sensing always outperform QST with incomplete information, unless measurements are performed on all 3-qubit Pauli bases. It turns out that QST fails to reconstruct TMI given incomplete Pauli observables.

We further consider the sampling noise and propose observables-based compressed sensing (CS Obsv.) by introducing binomial proportion confidence intervals (CI) into the expectations of observables, \textit{i.e.}, 
\begin{equation}
\begin{aligned}
    &\operatorname*{min}\left|\left|\sigma\right|\right|_{\mathrm{Tr}}\,\,\mathrm{s.t.}\,\,\mathrm{Tr}\left(\sigma\right)=1,\\
    &\mathrm{Tr}\left(\sigma P_k\right)\in\mathrm{CI}\left(\mathrm{Tr}\left(\rho P_k\right);m,1-\alpha\right),
\end{aligned}
\label{eq:Eq12}
\end{equation}
where $m$ is the number of samples for observable $P_k$. We linearly transform the measuring result of observables into $\frac 12 \left(\mathrm{Tr}\left(\rho P_k\right)+1\right)$, which is interpreted as a binomial proportion. The proportion, together with $m$, leads to the construction of a CI, which $\mathrm{Tr}\left(\sigma P_k\right)$ has probability $1-\alpha$ to fall into. 
Fine-tuning results show that Jeffreys intervals~\cite{brown2001interval} with $\alpha=1/3$ yields best performance in predicting TMI.

We compare the performance of the proposed CS Obsv., sampling-based compressed sensing (CS Samp.)~\cite{steffens2017experimentally} and machine learning (ML). 
Unlike our approach of using confidence intervals, CS Samp. extends CS to the noisy regime by constraining the samples of $\sigma$ in different basis with respect to experimental samples~\cite{steffens2017experimentally}. Machine learning also proves to be effective in predicting many-body properties~\cite{torlai2018neural}. As shown in Fig.~\ref{fig:Fig4} (b), given various number of samples, CS Obsv. constantly yields best performance, with MSE effectively decreasing as number of samples increases. On the contrary, ML and CS Samp. performances fail to improve steadily with number of samples, surpassed by QST given higher number of samples. 

\emph{Conclusions.}--In this work, we propose an on-chip simulation of an open quantum system that is able to accurately investigate the effects of system-environment interactions and memory effects of the environment on information scrambling. Our scheme fully utilizes the ability of the photonic chip to build cascaded MZIs on a large scale and precisely tune the collision matrix. Although each collision requires access to a fresh experimental mode, distributed quantum computing, which has been exploding recently, can be combined with our scheme to simulate larger-scale open quantum systems. Based on our conclusions that strong interaction between system and environment leads to a fast scrambling and large non-Markovianity causes persistent oscillations of information flow, future research may map specific thermodynamic processes by decomposing the evolution equation, e.g., the Lindblad master equation, for on-chip photonic simulations.


\section*{Acknowledgments} 
The authors thank M. S. Kim, Adam Taylor, Gabriela Bressanini, Yuxuan Zhang and Zhe-Yong Zhang for helpful discussions. 
This research is supported by the National Key R\&D Program of China (Grants No. 2024YFA1409300, No. 2019YFA0308703, No. 2019YFA0706302, and No. 2017YFA0303700); National Natural Science Foundation of China (NSFC) (Grants No. 62235012, No. 11904299, No. 61734005, No. 11761141014, and No. 11690033, No. 12104299, and No. 12304342); Innovation Program for Quantum Science and Technology (Grants No. 2021ZD0301500, and No. 2021ZD0300700); Science and Technology Commission of Shanghai Municipality (STCSM) (Grants No. 20JC1416300, No. 2019SHZDZX01, No.21ZR1432800, No. 22QA1404600, No. 24ZR1438700, No. 24ZR1430700 and No. 24LZ1401500); Shanghai Municipal Education Commission (SMEC) (Grants No. 2017-01-07-00-02-E00049); China Postdoctoral Science Foundation (Grants No. 2020M671091, No. 2021M692094, No. 2022T150415); Startup Fund for Young Faculty at SJTU (SFYF at SJTU)(Grants No. 24X010502876 and No. 24X010500170). X.-M.J. acknowledges additional support from a Shanghai talent program and support from Zhiyuan Innovative Research Center of Shanghai Jiao Tong University. H. T. acknowledges additional support from Yangyang Development Fund.

\section*{Competing interests}
The authors declare no competing interests.


\begin{thebibliography}{99}


\bibitem{von1950theory}
\bibinfo{author}{Von Bertalanffy, L.}
\bibinfo{title}{The theory of open systems in physics and biology.}
\newblock \emph{\bibinfo{journal}{Science}}
\textbf{\bibinfo{volume}{111}}, \bibinfo{pages}{23-29}
  (\bibinfo{year}{1950}).
  
\bibitem{diehl2008quantum}
\bibinfo{author}{Diehl, S., Micheli, A., Kantian, A., Kraus, B., Büchler, H. P. \& Zoller, P.}
\bibinfo{title}{Quantum states and phases in driven open quantum systems with cold atoms.}
\newblock \emph{\bibinfo{journal}{Nature Physics}}
\textbf{\bibinfo{volume}{4}}, \bibinfo{pages}{878-883}
  (\bibinfo{year}{2008}). 

\bibitem{krauter2011entanglement}
\bibinfo{author}{Krauter, H. \textit{et al.}}
\bibinfo{title}{Entanglement generated by dissipation and steady state entanglement of two macroscopic objects.}
\newblock \emph{\bibinfo{journal}{Physical Review Letters}}
\textbf{\bibinfo{volume}{107}}, \bibinfo{pages}{080503}
  (\bibinfo{year}{2011}). 

\bibitem{cho2011optical}
\bibinfo{author}{Cho, J., Bose, S. \& Kim, M. S.}
\bibinfo{title}{Optical pumping into many-body entanglement.}
\newblock \emph{\bibinfo{journal}{Physical Review Letters}}
\textbf{\bibinfo{volume}{106}}, \bibinfo{pages}{020504}
  (\bibinfo{year}{2011}). 
  
\bibitem{verstraete2009quantum}
\bibinfo{author}{Verstraete, F., Wolf, M. M. \& Cirac, J. I.}
\bibinfo{title}{Quantum computation and quantum-state engineering driven by dissipation.}
\newblock \emph{\bibinfo{journal}{Nature Physics}}
\textbf{\bibinfo{volume}{5}}, \bibinfo{pages}{633-636}
  (\bibinfo{year}{2009}).

\bibitem{giovannetti2011advances}
\bibinfo{author}{Giovannetti, V., Lloyd, S. \& Maccone, L.}
\bibinfo{title}{Advances in quantum metrology.}
\newblock \emph{\bibinfo{journal}{Nature Photonics}}
\textbf{\bibinfo{volume}{5}}, \bibinfo{pages}{222-229}
  (\bibinfo{year}{2011}).

\bibitem{giovannetti2006quantum}
\bibinfo{author}{Giovannetti, V., Lloyd, S. \& Maccone, L.}
\bibinfo{title}{Quantum metrology.}
\newblock \emph{\bibinfo{journal}{Physical Review Letters}}
\textbf{\bibinfo{volume}{96}}, \bibinfo{pages}{010401}
  (\bibinfo{year}{2006}).
  
\bibitem{gorini1976completely}
\bibinfo{author}{Gorini, V., Kossakowski, A. \& Sudarshan, E. C. G.}
\bibinfo{title}{Completely positive dynamical semigroups of N-level systems.}
\newblock \emph{\bibinfo{journal}{Journal of Mathematical Physics}}
\textbf{\bibinfo{volume}{17}}, \bibinfo{pages}{821-825}
  (\bibinfo{year}{1976}).

\bibitem{lindblad1976generators}
\bibinfo{author}{Lindblad, G.}
\bibinfo{title}{On the generators of quantum dynamical semigroups.}
\newblock \emph{\bibinfo{journal}{Communications in Mathematical Physics}}
\textbf{\bibinfo{volume}{48}}, \bibinfo{pages}{119-130}
  (\bibinfo{year}{1976}).  

\bibitem{carmichael1993quantum}
\bibinfo{author}{Carmichael, H. J.}
\bibinfo{title}{Quantum trajectory theory for cascaded open systems.}
\newblock \emph{\bibinfo{journal}{Physical Review Letters}}
\textbf{\bibinfo{volume}{70}}, \bibinfo{pages}{2273}
  (\bibinfo{year}{1993}).
  
  

  
\bibitem{dum1992monte}
\bibinfo{author}{Dum, R., Zoller, P. \& Ritsch, H.}
\bibinfo{title}{Monte Carlo simulation of the atomic master equation for spontaneous emission.}
\newblock \emph{\bibinfo{journal}{Physical Review A}}
\textbf{\bibinfo{volume}{45}}, \bibinfo{pages}{4879}
  (\bibinfo{year}{1992}). 
  
  
  \bibitem{Breuer2016}
 \bibinfo{author}{Breuer, H.-P., Laine, E.-M., Piilo, J. \& Vacchini, B.}
 \bibinfo{title}{Colloquium: Non-Markovian dynamics in open quantum systems.}
 \newblock \emph{\bibinfo{journal}{Reviews of Modern Physics}}
 \textbf{\bibinfo{volume}{88}}, \bibinfo{pages}{021002}
   (\bibinfo{year}{2016}).
  
  
  
  
  
   \bibitem{Swingle2016}
 \bibinfo{author}{Swingle, B., Bentsen, G., Schleier-Smith, M. \& Hayden, P.}
 \bibinfo{title}{Measuring the scrambling of quantum information.}
 \newblock \emph{\bibinfo{journal}{Physical Review A}}
 \textbf{\bibinfo{volume}{94}}, \bibinfo{pages}{040302}
   (\bibinfo{year}{2016}).
  
  \bibitem{Bohrdt2017}
\bibinfo{author}{Bohrdt, A., Mendl, C. B., Endres, M. \& Knap, M.}
\bibinfo{title}{Scrambling and thermalization in a diffusive quantum many-body system.}
\newblock \emph{\bibinfo{journal}{New Journal of Physics}}
\textbf{\bibinfo{volume}{19}}, \bibinfo{pages}{063001}
  (\bibinfo{year}{2017}).
  
  \bibitem{Campisi2017}
\bibinfo{author}{Campisi, M. \& Goold, J.}
\bibinfo{title}{Thermodynamics of quantum information scrambling.}
\newblock \emph{\bibinfo{journal}{Physical Review E}}
\textbf{\bibinfo{volume}{95}}, \bibinfo{pages}{062127}
  (\bibinfo{year}{2017}).
  
  \bibitem{Fan2017}
\bibinfo{author}{Fan, R., Zhang, P., Shen, H. \& Zhai, H.}
\bibinfo{title}{Out-of-time-order correlation for many-body localization.}
\newblock \emph{\bibinfo{journal}{Science Bulletin}}
\textbf{\bibinfo{volume}{62}}, \bibinfo{pages}{707-711}
  (\bibinfo{year}{2017}).
  
  \bibitem{Swingle2017}
\bibinfo{author}{Swingle, B. \& Chowdhury, D.}
\bibinfo{title}{Slow scrambling in disordered quantum systems.}
\newblock \emph{\bibinfo{journal}{Physical Review B}}
\textbf{\bibinfo{volume}{95}}, \bibinfo{pages}{060201}
  (\bibinfo{year}{2017}).
  
  \bibitem{Swingle2018}
\bibinfo{author}{Swingle, B. \& Yunger Halpern, N.}
\bibinfo{title}{Resilience of scrambling measurements.}
\newblock \emph{\bibinfo{journal}{Physical Review A}}
\textbf{\bibinfo{volume}{97}}, \bibinfo{pages}{062113}
  (\bibinfo{year}{2018}).
  
  \bibitem{Gonzalez2019}
\bibinfo{author}{Gonz$\acute{\mathrm{a}}$lez Alonso, J. R., Yunger Halpern, N. \& Dressel, J.}
\bibinfo{title}{Out-of-Time-Ordered-Correlator Quasiprobabilities Robustly Witness Scrambling.}
\newblock \emph{\bibinfo{journal}{Physical Review Letters}}
\textbf{\bibinfo{volume}{122}}, \bibinfo{pages}{040404}
  (\bibinfo{year}{2019}).
  
   \bibitem{Landsman2019}
 \bibinfo{author}{Landsman, K. A. \textit{et al.}}
 \bibinfo{title}{Verified quantum information scrambling.}
 \newblock \emph{\bibinfo{journal}{Nature}}
 \textbf{\bibinfo{volume}{567}}, \bibinfo{pages}{61-65}
   (\bibinfo{year}{2019}).

  \bibitem{Yoshida2019}
\bibinfo{author}{Yoshida, B. \& Yao, N. Y.}
\bibinfo{title}{Disentangling Scrambling and Decoherence via Quantum Teleportation.}
\newblock \emph{\bibinfo{journal}{Physical Review X}}
\textbf{\bibinfo{volume}{9}}, \bibinfo{pages}{011006}
  (\bibinfo{year}{2019}).

\bibitem{Joshi2020Quantum}
\bibinfo{author}{Joshi, M. K., Elben, A., Vermersch, B., Brydges, T., Maier, C., Zoller, P., Blatt, R. \& Roos, C. F.}
\bibinfo{title}{Quantum information scrambling in a trapped-ion quantum simulator with tunable range interactions.}
\newblock \emph{\bibinfo{journal}{Physical Review Letters}}
\textbf{\bibinfo{volume}{124}}, \bibinfo{pages}{240505}
(\bibinfo{year}{2020}).

  \bibitem{Mi2021}
\bibinfo{author}{Mi, X. \textit{et al.}}
\bibinfo{title}{Information scrambling in quantum circuits.}
\newblock \emph{\bibinfo{journal}{Science}}
\textbf{\bibinfo{volume}{374}}, \bibinfo{pages}{1479-1483}
  (\bibinfo{year}{2021}).


  \bibitem{Shen2020}
\bibinfo{author}{Shen, H., Zhang, P., You, Y.-Z. \& Zhai, H.}
\bibinfo{title}{Information Scrambling in Quantum Neural Networks.}
\newblock \emph{\bibinfo{journal}{Physical Review Letters}}
\textbf{\bibinfo{volume}{124}}, \bibinfo{pages}{200504}
  (\bibinfo{year}{2020}).

  \bibitem{Lin2021}
\bibinfo{author}{Lin, J.-D. \textit{et al.}}
\bibinfo{title}{Quantum steering as a witness of quantum scrambling.}
\newblock \emph{\bibinfo{journal}{Physical Review A}}
\textbf{\bibinfo{volume}{104}}, \bibinfo{pages}{022614}
  (\bibinfo{year}{2021}).
  
    \bibitem{Jafferis2022}
\bibinfo{author}{Jafferis, D. \textit{et al.}}
\bibinfo{title}{Traversable wormhole dynamics on a quantum processor.}
\newblock \emph{\bibinfo{journal}{Nature}}
\textbf{\bibinfo{volume}{612}}, \bibinfo{pages}{51-55}
  (\bibinfo{year}{2022}).

 \bibitem{Zhu2022}
\bibinfo{author}{Zhu, Q. \textit{et al.}}
\bibinfo{title}{Observation of Thermalization and Information Scrambling in a Superconducting Quantum Processor.}
\newblock \emph{\bibinfo{journal}{Physical Review Letters}}
\textbf{\bibinfo{volume}{128}}, \bibinfo{pages}{160502}
  (\bibinfo{year}{2022}).

  
  \bibitem{Hosur2016}
 \bibinfo{author}{Hosur, P., Qi, X.-L., Roberts, D. A. \& Yoshida, B.}
 \bibinfo{title}{Chaos in quantum channels.}
 \newblock \emph{\bibinfo{journal}{Journal of High Energy Physics}}
 \textbf{\bibinfo{volume}{2016}}, \bibinfo{pages}{4}
   (\bibinfo{year}{2016}).

  
  \bibitem{Bhattacharyya2022}
\bibinfo{author}{Bhattacharyya, A., Joshi, L. K. \& Sundar, B.}
\bibinfo{title}{Quantum information scrambling: from holography to quantum simulators.}
\newblock \emph{\bibinfo{journal}{The European Physical Journal C}}
\textbf{\bibinfo{volume}{82}}, \bibinfo{pages}{458}
  (\bibinfo{year}{2022}).




  \bibitem{Blok2021}
\bibinfo{author}{Blok, M. S. \textit{et al.}}
\bibinfo{title}{Quantum Information Scrambling on a Superconducting Qutrit Processor.}
\newblock \emph{\bibinfo{journal}{Physical Review X}}
\textbf{\bibinfo{volume}{11}}, \bibinfo{pages}{021010}
  (\bibinfo{year}{2021}).

  
 

  \bibitem{Bluvstein2024}
\bibinfo{author}{Bluvstein, D. \textit{et al.}}
\bibinfo{title}{Logical quantum processor based on reconfigurable atom arrays.}
\newblock \emph{\bibinfo{journal}{Nature}}
\textbf{\bibinfo{volume}{628}}, \bibinfo{pages}{58-65}
  (\bibinfo{year}{2024}).





  \bibitem{Li2020}
\bibinfo{author}{Li, Y., Li, X. \& Jin, J.}
\bibinfo{title}{Information scrambling in a collision model.}
\newblock \emph{\bibinfo{journal}{Physical Review A}}
\textbf{\bibinfo{volume}{101}}, \bibinfo{pages}{042324}
  (\bibinfo{year}{2020}).
  
  \bibitem{Li2022}
\bibinfo{author}{Li, Y., Li, X. \& Jin, J.}
\bibinfo{title}{Dissipation-Induced Information Scrambling in a Collision Model.}
\newblock \emph{\bibinfo{journal}{Entropy}}
\textbf{\bibinfo{volume}{24}}, \bibinfo{pages}{345}
  (\bibinfo{year}{2022}).

  \bibitem{Han2022}
\bibinfo{author}{Han, L.-P., Zou, J., Li, H. \& Shao, B.}
\bibinfo{title}{Quantum Information Scrambling in Non-Markovian Open Quantum System.}
\newblock \emph{\bibinfo{journal}{Entropy}}
\textbf{\bibinfo{volume}{24}}, \bibinfo{pages}{1532}
  (\bibinfo{year}{2022}).


\bibitem{Aspuru-Guzik2012}
\bibinfo{author}{Aspuru-Guzik, A. \& Walther, P.}
\bibinfo{title}{Photonic quantum simulation.}
\newblock \emph{\bibinfo{journal}{Nature Physics}}
\textbf{\bibinfo{volume}{8}}, \bibinfo{pages}{285-291}
  (\bibinfo{year}{2012}).
  
  
 \bibitem{Ralph2002}
 \bibinfo{author}{Ralph, T. C., Langford, N. K., Bell, T. B. \& White, A. G.}
 \bibinfo{title}{Linear optical controlled-NOT gate in the coincidence basis.}
 \newblock \emph{\bibinfo{journal}{Physical Review A}}
 \textbf{\bibinfo{volume}{65}}, \bibinfo{pages}{062324}
   (\bibinfo{year}{2002}).

  
  
  
  
  
  
  
  
%


  \bibitem{Ciccarello2013}
\bibinfo{author}{Ciccarello, F., Palma, G. M. \& Giovannetti, V.}
\bibinfo{title}{Collision-model-based approach to non-Markovian quantum dynamics.}
\newblock \emph{\bibinfo{journal}{Physical Review A}}
\textbf{\bibinfo{volume}{87}}, \bibinfo{pages}{040103}
  (\bibinfo{year}{2013}).
  
  \bibitem{McCloskey2014}
\bibinfo{author}{McCloskey, R. \& Paternostro, M.}
\bibinfo{title}{Non-Markovianity and system-environment correlations in a microscopic collision model.}
\newblock \emph{\bibinfo{journal}{Physical Review A}}
\textbf{\bibinfo{volume}{89}}, \bibinfo{pages}{052120}
  (\bibinfo{year}{2014}).
  
  \bibitem{Ciccarello2017}
\bibinfo{author}{Ciccarello, F.}
\bibinfo{title}{Collision models in quantum optics.}
\newblock \emph{\bibinfo{journal}{Quantum Measurements and Quantum Metrology}}
\textbf{\bibinfo{volume}{4}}, \bibinfo{pages}{53-63}
  (\bibinfo{year}{2017}).
  
  
  \bibitem{Jin2018}
\bibinfo{author}{Jin, J. \& Yu, C.}
\bibinfo{title}{Non-Markovianity in the collision model with environmental block.}
\newblock \emph{\bibinfo{journal}{New Journal of Physics}}
\textbf{\bibinfo{volume}{20}}, \bibinfo{pages}{053026}
  (\bibinfo{year}{2018}).
  
  \bibitem{Cuevas2019}
\bibinfo{author}{Cuevas, A. \textit{et al.}}
\bibinfo{title}{All-optical implementation of collision-based evolutions of open quantum systems.}
\newblock \emph{\bibinfo{journal}{Scientific Reports}}
\textbf{\bibinfo{volume}{9}}, \bibinfo{pages}{3205}
  (\bibinfo{year}{2019}).
  
  \bibitem{Rodrigues2019}
\bibinfo{author}{Rodrigues, F. L. S., De Chiara, G., Paternostro, M. \& Landi, G. T.}
\bibinfo{title}{Thermodynamics of Weakly Coherent Collisional Models.}
\newblock \emph{\bibinfo{journal}{Physical Review Letters}}
\textbf{\bibinfo{volume}{123}}, \bibinfo{pages}{140601}
  (\bibinfo{year}{2019}).
  
  
  \bibitem{Garcia2020}
\bibinfo{author}{Garcia-P\'erez, G., Rossi, M. A. C. \& Maniscalco, S.}
\bibinfo{title}{IBM Q Experience as a versatile experimental testbed for simulating open quantum systems.}
\newblock \emph{\bibinfo{journal}{npj Quantum Information}}
\textbf{\bibinfo{volume}{6}}, \bibinfo{pages}{1}
  (\bibinfo{year}{2020}).
  
  
  \bibitem{Cattaneo2021}
\bibinfo{author}{Cattaneo, M., De Chiara, G., Maniscalco, S., Zambrini, R. \& Giorgi, G. L.}
\bibinfo{title}{Collision Models Can Efficiently Simulate Any Multipartite Markovian Quantum Dynamics.}
\newblock \emph{\bibinfo{journal}{Physical Review Letters}}
\textbf{\bibinfo{volume}{126}}, \bibinfo{pages}{130403}
  (\bibinfo{year}{2021}).
  
  
  \bibitem{Ciccarello2022}
\bibinfo{author}{Ciccarello, F., Lorenzo, S., Giovannetti, V. \& Palma, G. M.}
\bibinfo{title}{Quantum collision models: Open system dynamics from repeated interactions.}
\newblock \emph{\bibinfo{journal}{Physics Reports}}
\textbf{\bibinfo{volume}{954}}, \bibinfo{pages}{1-70}
  (\bibinfo{year}{2022}).
  
  \bibitem{Cusumano2022}
\bibinfo{author}{Cusumano, S.}
\bibinfo{title}{Quantum Collision Models: A Beginner Guide.}
\newblock \emph{\bibinfo{journal}{Entropy}}
\textbf{\bibinfo{volume}{24}}, \bibinfo{pages}{1258}
  (\bibinfo{year}{2022}).
  
%
%
%


  
  \bibitem{Li2024}
\bibinfo{author}{Li, Y., Li, X. \& Jin, J.}
\bibinfo{title}{Witnessing non-Markovianity with Gaussian quantum steering in a collision model.}
\newblock \emph{\bibinfo{journal}{Physical Review A}}
\textbf{\bibinfo{volume}{109}}, \bibinfo{pages}{052201}
  (\bibinfo{year}{2024}).




\bibitem{gross2010quantum} 
\bibinfo{author}{Gross, D., Liu, Y.-K., Flammia, S. T., Becker, S. \& Eisert, J.}
\bibinfo{title}{Quantum state tomography via compressed sensing.} 
\newblock \emph{\bibinfo{journal}{Physical Review Letters}} 
\textbf{\bibinfo{volume}{105}}, \bibinfo{pages}{150401} (\bibinfo{year}{2010}).

  \bibitem{Kudler-Flam2020}
 \bibinfo{author}{Kudler-Flam, J., Nozaki, M., Ryu, S. \& Tan, M. T.}
 \bibinfo{title}{Quantum vs. classical information: operator negativity as a probe of scrambling.}
 \newblock \emph{\bibinfo{journal}{Journal of High Energy Physics}}
 \textbf{\bibinfo{volume}{2020}}, \bibinfo{pages}{31}
   (\bibinfo{year}{2020}).

\bibitem{Liu2011}
 \bibinfo{author}{Liu, B.-H. \textit{et al.}}
 \bibinfo{title}{Experimental control of the transition from Markovian to non-Markovian dynamics of open quantum systems.}
 \newblock \emph{\bibinfo{journal}{Nature Physics}}
 \textbf{\bibinfo{volume}{7}}, \bibinfo{pages}{931-934}
   (\bibinfo{year}{2011}).





  \bibitem{McGinley2022}
\bibinfo{author}{McGinley, M., Leontica, S., Garratt, S. J., Jovanovic, J. \& Simon, S. H.}
\bibinfo{title}{Quantifying information scrambling via classical shadow tomography on programmable quantum simulators.}
\newblock \emph{\bibinfo{journal}{Physical Review A}}
\textbf{\bibinfo{volume}{106}}, \bibinfo{pages}{012441}
  (\bibinfo{year}{2022}).

\bibitem{riofrio2017experimental} 
\bibinfo{author}{Riofrio, C. A., Gross, D., Flammia, S. T., Monz, T., Nigg, D., Blatt, R. \& Eisert, J.} 
\bibinfo{title}{Experimental quantum compressed sensing for a seven-qubit system.}
\newblock \emph{\bibinfo{journal}{Nature Communications}} 
\textbf{\bibinfo{volume}{8}}, \bibinfo{pages}{15305} (\bibinfo{year}{2017}).


\bibitem{steffens2017experimentally}
\bibinfo{author}{Steffens, A., Riofrio, C. A., McCutcheon, W., Roth, I., Bell, B. A., McMillan, A., Tame, M. S., Rarity, J. G. \& Eisert, J.}
\bibinfo{title}{Experimentally exploring compressed sensing quantum tomography.}
\newblock \emph{\bibinfo{journal}{Quantum Science and Technology}}
\textbf{\bibinfo{volume}{2}}, \bibinfo{pages}{025005}
  (\bibinfo{year}{2017}).
  
\bibitem{brown2001interval} 
\bibinfo{author}{Brown, L. D., Cai, T. T. \& DasGupta, A.}
\bibinfo{title}{Interval estimation for a binomial proportion.}
\newblock {\emph{\bibinfo{journal}{Statistical Science}}}
\textbf{\bibinfo{volume}{16}}, \bibinfo{pages}{101--133} (\bibinfo{year}{2001}).

\bibitem{torlai2018neural}
\bibinfo{author}{Torlai, G., Mazzola, G., Carrasquilla, J., Troyer, M., Melko, R. \& Carleo, G.}
\bibinfo{title}{Neural-network quantum state tomography.}
\newblock \emph{\bibinfo{journal}{Nature Physics}}
\textbf{\bibinfo{volume}{14}}, \bibinfo{pages}{447--450}
  (\bibinfo{year}{2018}).
  

%

  

\end{thebibliography}
\end{document}